\documentclass[aps,prl,reprint,superscriptaddress,nobalancelastpage,twocolumn,showpacs]{revtex4-1}
\usepackage{color,amsthm,amsmath,amsfonts,graphicx,bm,amssymb, tikz, braket, comment}
\usepackage{epstopdf}
\usepackage{pdfpages}
\makeatletter
\AtBeginDocument{\let\LS@rot\@undefined}
\makeatother

\usetikzlibrary{calc}
\usepackage{etoolbox}
\usepackage{marginnote}
\begin{document}
\newcommand{\beq}{\begin{equation}}
\newcommand{\eeq}{\end{equation}}
\newcommand{\Ord}[1]{{O}\left(#1\right)}
\newcommand{\sign}{\,{\rm sign}}
\newcommand{\arctanh}{\,{\rm arctanh}}
\newcommand{\subscript}[2]{$#1 _ #2$}

\providecommand{\abs}[1]{\lvert#1\rvert}  
\providecommand{\norm}[1]{\lVert#1\rVert}

\newcommand{\twovec}[2]{\left[\begin{array}{c}
#1\\#2
\end{array}
\right]}

\newcommand{\avg}[2]{\langle #1 \rangle_{\mbox{\tiny #2}}}

\newcommand\makered[1]{{\color{red} #1}}
\newcommand\makeblue[1]{{\color{blue} #1}}
\newcommand\repl[2]{\sout{#1}\makered{#2}}
\newcommand\mrepl[2]{\text{\sout{\ensuremath{#1}}}\makered{#2}}
\newcommand\note[1]{\marginnote{\tiny \textcolor{blue}{#1}}}

\newcommand\leftnote[1]{\reversemarginpar\marginnote{\tiny \textcolor{blue}{#1}}}
\newcommand\rightnote[1]{\normalmarginpar\marginnote{\tiny \textcolor{blue}{#1}}}
\def\vmax{v_{\mbox{\tiny max}}}

\title{Exact logarithmic four-point functions in the critical two-dimensional Ising model}

\author{Giacomo Gori}
\affiliation{SISSA \& CNR-IOM, Via Bonomea 265, 34136 Trieste, Italy}
\author{Jacopo Viti}
\affiliation{ECT  \& Instituto Internacional de F\'isica, UFRN, Lagoa Nova 59078-970 Natal, Brazil}
\thanks{\textit{E-mail for correspondence:} \texttt{viti.jacopo@gmail.com} }

\begin{abstract}
Based on conformal symmetry we propose an exact formula for the four-point connectivities of 
FK clusters in the critical Ising model when the four points are anchored to the boundary. The explicit
solution we found displays logarithmic singularities. We check our prediction using Monte Carlo simulations on
a triangular lattice, showing excellent agreement. Our findings could shed further light on the formidable task
of the characterization of Logarithmic Conformal Field Theories and on their relevance in physics.

\end{abstract}



\maketitle

\paragraph*{Introduction --} 
Conformal symmetry in two dimensions~\cite{BPZ}  has been of extraordinary usefulness 
to study classical statistical mechanics models at criticality since the 80's. It has 
notably found also extensive applications in the quantum realm, spanning from 
gapless one-dimensional systems~\cite{Haldane}, quantum Hall effect~\cite{MR} and  
entanglement~\cite{Entanglement}. Two dimensional Conformal Invariant Quantum Field 
Theories (CFTs) and in particular Liouville theory are moreover the cornerstone of 
world-sheet geometry in string theory~\cite{Polyakov}. The simplest CFTs that capture 
the critical behaviour of lattice models and quantum spin chains are \textit{unitary}. Moreover, when their Hilbert space splits into a  finite number of
representations of the conformal (or some larger) symmetry are usually termed \textit{rational}. CFTs are classified according to their central charge $c$; for instance for a free massless boson $c=1$.

However unitary and rational theories are far from exhausting the physically
relevant conformally invariant theories.
In the beginning of the 90's the groundbreaking
discovery~\cite{Cardy_perc} of an exact formula for the crossing probability 
in critical percolation forced to analyze theories violating these two assumptions. Percolation is a simple stochastic process where bonds or sites on a lattice can be occupied independently with a certain probability. 
Since the partition function is not affected by finite-size effects,  the central charge of a putative CFT describing critical percolation is zero~\cite{BCM}. The existence of a non-trivial 
formula for the crossing probability makes it a prominent example of a non-unitary CFT (the only unitary CFT with $c=0$ is trivial).
For subsequent developments leading to the formulation of the Stochastic Lowener Evolution we refer to the reviews~\cite{BB, Cardy_SLE}.
 At the same time,  it was  recognized that as 
far as the conditions of unitarity and rationality are relaxed, CFT correlation 
functions can display striking logarithmic singularities that are actually the 
signatures of intricate realizations of the conformal symmetry~\cite{RSal, G}.  The class 
of non-unitary and generally non-rational CFTs where these unconventional features 
show up, was christened \textit{Logarithmic CFTs}. Such theories were promptly 
argued to play a fundamental role in the characterization of disordered systems 
in two dimensions~\cite{Caux, Cardy_99, GL}, for instance by means of 
supersymmetry~\cite{GLR, RS}.  

With these motivations, Logarithmic CFTs  were investigated in greater detail in 
the last decade, either from a purely algebraic point of view~\cite{Ridout_rev}, either constructing lattice regularizations~\cite{RS, Logmin, RS1, DJS}
or generalizing the study of crossing formulas in critical 
percolation~\cite{Watts, Simmons, Flores}.  Recently~\cite{DV}, it has also been suggested that an analytic non-unitary 
extension of  Liouville theory~\cite{Zam} might describe the connectivity 
properties of critical bulk percolation and more generally the $Q$-state Potts 
model. The domain of applicability of Liouville theory in statistical mechanics is currently an important open problem see~\cite{IJS, PRS} and also~\cite{RS15}. 

Despite these huge advances, a satisfactory understanding of Logarithmic CFTs is still a long way off.
Moreover, it is fair to say that  few examples of explicit 
logarithmic singularities have been found in familiar statistical 
models: notably only in percolation~\cite{VJS, FSK2}
($c=0$), dense polymers~\cite{Sal_pol, GK, K00}; see also~\cite{Santa2, Vasseur_2016} for  applications to 
disordered systems. An exact Coulomb gas approach  to CFT correlation functions~\cite{KF, KF_1,KF_2} closely related to those 
considered in this letter reveals an infinite number of
logarithmic cases. These arise from operator mixing as is the case here. These results  suggest the possibility of logarithmic
behavior in multiple Self Avoiding Random Walks (SAWs)~\cite{KF}.

In this paper we show  how a logarithmic singularity due to operator mixing also arises in the 
context of arguably the best-known model of statistical mechanics: the 
two-dimensional Ising model; see~\cite{Aldo_rev, Giuseppe_book} for a pedagogical 
introduction  to the richness of the model. This is remarkable since it shows unambiguously that critical properties of Ising clusters are ruled by a Logarithmic
CFT.  Moreover the four-point function we consider here, a four point 
connectivity in the Fortuin and Kasteleyn (FK) representation of the Ising 
model, is a natural observable that can be easily simulated with Monte Carlo (MC)
methods.  Logarithmic singularities in Ising connectivities were  also
argued to exist in~\cite{Cardy_revLog, VJ_log}, in this letter we
demonstrate it explicitly using CFT.

 Finally,
our findings could shed further light on the extremely challenging 
problem of the characterization of Logarithmic CFTs and on their applications to physics.

\begin{figure}[t]
\centering
\includegraphics[scale=1]{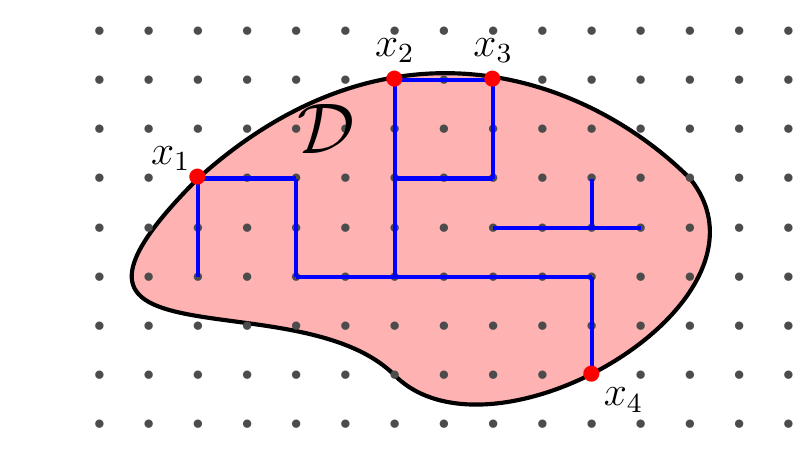}
\caption{The $Q$-state Potts model is defined on a finite domain $\mathcal{D}$ of
the plane. FK graphs $G$ on the underlying square lattice are drawn in blue. The figure shows a particular configuration
where the four boundary points $x_1,~x_2,~x_3$ and $x_4$ are all connected
by an FK cluster thus
contributing to $P_{(1234)}$.}
\label{fig:FK}
\end{figure}

\paragraph*{Four points connectivities in the  Ising model --}
It is convenient to introduce the Ising FK clusters,
starting from the ferromagnetic $Q$-state Potts model~\cite{Wu};  The Ising case is recovered by setting $Q=2$.
The model is defined on a finite simply-connected domain $\mathcal{D}$ of the plane,
see Fig.~\ref{fig:FK}, and  the choice of the underlying lattice
is  irrelevant at the critical point.
To introduce the notion of cluster connectivities, we
should first recall the Fortuin-Kastelein (FK) representation~\cite{FK}. The $Q$-state 
Potts  model is defined in terms of spin variables $s(x)$ taking $1,\dots, Q$ different values; its partition function can be expressed as a product
 over the lattice edges as
 \begin{equation}
 \label{pf}
 Z=\sum_{\{s(x)\}}\prod_{\langle x,y\rangle}[(1-p)+p\delta_{s(x),s(y)}],
 \end{equation}
where $p$ is a parameter related to the temperature and the product in \eqref{pf}
extends only to next-neighboring sites.
Suppose then to expand such a product: Each term in the expansion can be 
represented graphically by drawing a bond between $x$ and $y$ if the factor 
$p\delta_{s(x)s(y)}$ is selected and leaving empty the bond if
such a factor is absent.
The set of non-empty bonds in each term of the expansion then defines  
a graph $G$ on the underlying lattice that is called
FK graph. Such a graph might contain $N_c$ different connected components 
(including isolated points), dubbed FK clusters.
Moreover, on each cluster the spin values 
are constrained to be the same because of the Kronecker delta in \eqref{pf}. 
Summing over their possible $Q$ values leads
to the following rewriting of the $Q$-state Potts model partition function as a sum over graphs: 
 $Z=\sum_{G} (1-p)^{\bar{n}_b}p^{n_b} Q^{N_c},$ where $n_b$ and $\bar{n}_b$ are the number of occupied and empty bonds in the graph $G$. 
For arbitrary non-negative  $Q$, the graph representation for $Z$ is a generalized
percolation problem known as Random Cluster Model, where bonds occupied with probability
$p$ are not independent random variables.
The fundamental observables in the Random Cluster Model are the
connectivities and they offer a purely geometrical
interpretation of the magnetic Potts model phase transition. Connectivities
represent the different probabilities with which
$n$ points of the plane can be partitioned into FK clusters. 
If the points are on the boundary
of the domain $\mathcal{D}$, the total number of $n$-point connectivities is clearly equal to
the number of non-crossing partitions of a set of $n$ elements,
i.e. the Catalan number $C_n$; for example
if $n=4$ there  are $C_4=14$ of  them. These  functions are
however 
not linearly independent, since they satisfy sum rules:
for instance the sum over all $n$-point connectivities has to be one.
Following~\cite{DVPotts}, it is possible to show that a valid choice of
$n$-point linearly
independent connectivities is given by all the probabilities associated to
configurations where no point is
disconnected from all the others (non-singleton partitions). In the specific example of $n=4$ and $x_1$, $x_2$, $x_3$, $x_4$ 
on the boundary of $\mathcal{D}$, see again Fig.~\ref{fig:FK}, a possible choice
of linearly independent connectivities  is: 
$P_{(1234)}$, $P_{(12)(34)}$ and $P_{(14)(23)}$. The function $P_{(1234)}$ denotes the probability that all
the four points $x_1$, $x_2$, $x_3$ and $x_4$ are on the same FK cluster;
$P_{(12)(34)}$  is instead the probability that $x_1$ and $x_2$ are in the same cluster,
$x_3$ and $x_4$ in the same cluster but these two are now different and analogously
for $P_{(14)(23)}$. We also omitted for simplicity the explicit spacial dependence. Notice  that when the points $x_1,~x_2,~x_3$ and $x_4$ are anchored to the boundary the function $P_{(13)(24)}$ does not appear since  two clusters cannot cross.

\begin{figure} 
\includegraphics[scale=1]{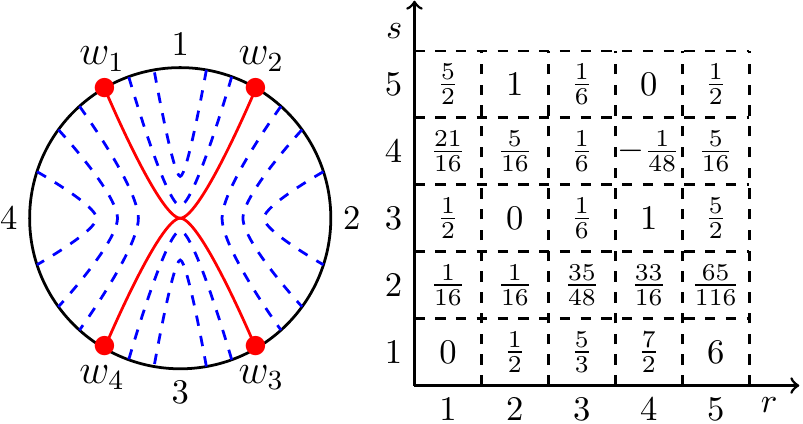}
\caption{On the left, schematic representation of allowed dual FK clusters
(dashed blue curves) in the Potts model when boundary conditions that fix the values of the
dual boundary spins  to $1,2,3$ and $4$ are chosen. On the right, the Kac table, obtained from the scaling dimension $h_{r,s}$ in \eqref{scaling} for $Q=2$ 
corresponding to the Ising model.}
\label{fig:conn}
\end{figure}

\paragraph*{Exact solution--} We 
 turn
now to the exact determination of these three functions in the critical Ising model,
using arguments inspired by
the seminal work~\cite{Cardy_perc}. At the critical point, 
conformal invariance allows one to map any simply connected domain $\mathcal{D}$ of the plane 
by the Riemann mapping theorem into the unit disk. Moreover, the points $x_1$, $x_2$, 
$x_3$, $x_4$ are mapped to points $w_1$, $w_2$, $w_3$, $w_4$ lying at the
boundary (circumference) of such a disk.  The three connectivities $P_{(1234)},~P_{(12)(34)}$ and $P_{(13)(24)}$
 can be singled out by computing Potts partition functions with specific boundary conditions for 
 the dual Potts spins~\cite{DVPotts}.
 As an example let us suppose to fix the values of the dual Potts spins at the boundary of the 
 disk to be $1,2,3$ and $4$ as in Fig.~\ref{fig:conn}
 left and to compute the Potts partition function in this case. Notice that this assignment will require at
 least four available colors, i.e. $Q\geq 4$, and it would be non-physical for the Ising model.
 It has however certainly sense if we assume $Q$ real and imagine to compute connectivities
 in the Random Cluster Model at any values of $Q$ and take eventually the limit $Q\rightarrow 2$. Configurations 
 of dual FK clusters with such a particular choice of boundary conditions
 cannot contain clusters that cross from regions with boundary conditions $\alpha$ to regions
 with boundary conditions $\beta$ if $\alpha\not=\beta$. Dual FK clusters are represented 
 schematically by blue dashed curves in Fig.~\ref{fig:conn}.
 Applying a duality transformation to the Potts model partition function~\cite{Wu},
 these configurations are in one-to-one correspondence with configurations
 where a single FK cluster, continuous red curve in Fig.~\ref{fig:conn},
 connects the four boundary points.  The reasoning above allows
 to compute then connectivities as Potts partition functions with insertion of
 local operators $\phi_{(\alpha|\beta)}$ that switch the values of the dual spins at
 the boundary from  $\alpha\rightarrow\beta$, $\alpha,\beta=1,\dots,Q$. In the jargon of
 CFT, the fields $\phi_{(\alpha|\beta)}$ are called boundary-condition-changing
 operators. In this way, we can argue for example that $P_{(1234)}$ has to be
 proportional to the correlation function of
 $\langle\phi_{(4|1)}(w_1)\phi_{(1|2)}(w_2)\phi_{(2|3)}(w_3)\phi_{(3|4)}(w_4)\rangle$.

 Let us briefly recall that in the simplest case,
the scaling fields $\phi_{r,s}$ of any CFT can be classified by two positive integers $r,s$ such
that their scaling dimensions are
\begin{equation}
\label{scaling}
 h_{r,s}=\frac{[r(m+1)-sm]^2-1}{4m(m+1)},~~m\in\mathbb R.
\end{equation}
The parameter $m$ is related to the central charge $c$ of the CFT through $c(m)=1-\frac{6}{m(m+1)}$, 
and in turns for the Potts model $Q=4\cos^2[\pi/(m+1)]$. The values $h_{r,s}$ can be represented into 
a lattice, dubbed the Kac table; for a CFT with $c=1/2$ as the Ising model, the Kac table is 
represented in Fig.~\ref{fig:conn} on the right.

 The boundary condition changing operator $\phi_{(\alpha|\beta)}$  was identified in~\cite{Cardy_perc} for any
values of $Q$ as the field $\phi_{1,3}$. Notice that at $c=1/2$, the dimension of $\phi_{1,3}$ is 
$h_{1,3}=1/2$ and coincides with the one of the Ising order parameter $\sigma$, when inserted at the 
boundary~\cite{Cardy_verlinde}; in this case the spin operator $\sigma$ transforms as the field~$\phi_{2,1}$. 
In the construction of the simplest  conformal field theory describing the $\mathbb{Z}_2$ universality class 
these two fields can  be actually  identified and consequently the operator product algebra of 
$\{\phi_{1,1},\phi_{2,1},{\phi_{1,2}}\}$ closes. The self-consistent closure of the operator product 
algebra was used as a criterion in~\cite{BPZ} to build the whole family of minimal conformal models, 
where only a finite numbers of Virasoro algebra representations should be considered and furthermore 
allows  to classify all the possible  conformal boundary conditions~\cite{Cardy_verlinde}.  However when analyzing 
 the connectivity properties of the Ising FK clusters, the identification of 
$\phi_{1,3}$ with $\phi_{2,1}$ is no longer possible. According to the general theory~\cite{BPZ}, 
see also~\cite{DMS}, the four-point function of $\phi_{1,3}$ satisfies a 
linear differential equation of degree $3$. If we map the unit disk to the upper half plane $\mathbb H$ 
and call $z_1,\dots,z_4$ the images on the real axis of the boundary points $w_1,\dots,w_4$ we have
\begin{equation}
\label{corr}
\langle\prod_{i=1}^4\phi_{1,3}(z_i)\rangle_{\mathbb H}=
\left[\frac{z_{42}z_{31}}{z_{21}z_{43}z_{32}z_{14}}\right]^{2h_{1,3}}F(\eta);
\end{equation}
where $z_{ij}=z_i-z_j$ and $\eta=\frac{z_{21}z_{43}}{z_{42}z_{31}}$ is the harmonic ratio ($0<\eta<1$). 
For the Ising model, the function $F(\eta)$ is the solution of the  differential equation~\cite{Suppl}
\begin{equation}
\label{diff}
[2\eta(1-\eta)]^2F^{'''}-3(1-\eta+\eta^2)F'+3(2\eta-1)F=0.
\end{equation}
The equation~\eqref{diff} has three linearly independent solutions $F_{1,1}(\eta)$, $F_{1,3}(\eta)$, $F_{1,5}(\eta)$. 
The behaviour for small $\eta$  of each of the functions  
$F_{r,s}$ is of the form $\eta^{h_{r,s}}$ and the exponent $h_{r,s}$ coincides with the scaling dimension of the 
field $\phi_{r,s}$ that is  produced in  the operator product algebra~\cite{BPZ}: 
$\phi_{1,3}\times\phi_{1,3}=\phi_{1,1}+\phi_{1,3}+\phi_{1,5}$.
Although there is not a general procedure to solve the differential equation~\eqref{diff}, 
we can proceed as follows. Firstly, we  observe that function $F_{1,1}(\eta)$  has to coincide apart from 
the prefactor in~\eqref{corr} with the four point function of the boundary spin $\sigma$ and such a 
function~\cite{BPZ} is the simple monodromy invariant~\cite{Foot} polynomial $1-\eta+\eta^2$. It is also 
easy to understand what $F_{1,1}$ should be in terms of connectivities. Since the four-point function of 
the boundary spin operator can be fully defined in the minimal Ising model, it has to correspond to the 
unique partition function that requires only two colors to be construct, namely 
$\langle\phi_{(2|1)}(w_1)\phi_{(1|2)}(w_2)\phi_{(2|1)}(w_3)\phi_{(1|2)}(w_4)\rangle$.
This in turn, 
 is proportional to the sum $P_{t}=P_{(1234)}+P_{(14)(23)}+P_{(12)(34)}$ 
of the three linearly independent connectivities.  Using the known solution $F_{1,1}(\eta)$ we can reduce 
the degree of the differential equation~\eqref{diff} by substituting
$F(\eta)=F_{1,1}(\eta)\int^{\eta}_0 d\eta' G(\eta')$. The function $G(\eta)$ is finally obtained 
through a rational pull-back of the Gauss hypergeometric function~\cite{key}, see also~\cite{Suppl}. 
One gets  two linearly independent solutions $G_{1,2}$ for $G(\eta)$, related by the transformation $\eta\rightarrow 1-\eta$: 
$G_1(\eta)=f(\eta)$ and $G_2(\eta)=f(1-\eta)$. The function $f(\eta)$ is
\begin{equation}
f(\eta)=\frac{p(\eta)E(\eta)+ q(\eta)K(\eta)}{\sqrt{(1-\eta) \eta}},
\end{equation}
with $K(\eta)$ and $E(\eta)$ the elliptic integrals of first and second kind respectively, 
and $p(\eta)$ and $q(\eta)$ rational functions of $\eta$~\cite{Suppl}. The behaviour 
for small $\eta$ finally fixes, up to an overall constant,
 $F_{1,5}(\eta)=F_{1,1}(\eta)\int^{\eta}_0 d\eta' f(\eta')$. The third linear independent solution to~\eqref{diff} can be chosen to be $F_{1,5}(1-\eta)$ that is actually a  linear
combination~\cite{note} of all the $F$'s.
Coming back to the connectivities we observe that in the limit $w_1\rightarrow w_2$, $P_{(14)(23)}$ 
contains configurations where two FK clusters are separated by a dual line. These configurations are realized by the insertion of the operator $\phi_{1,5}$~\cite{Cardy_inc}
at the boundary and it was  argued in~\cite{Cardy_revLog} that in this case logarithmic singularities should arise. 
We conjecture then  the following identification for the universal probability ratio which we denote with $R(\eta)$
\begin{equation}
\label{ratio}
R(\eta)=\frac{P_{(14)(23)}(\eta)}{P_t(\eta)}=A\int_{0}^\eta d\eta' f(\eta')
\end{equation}  
where the constant $A=[\int_{0}^1 d\eta f(\eta)]^{-1}$ is chosen to ensure that 
$R(1)=1$.
 The conjecture~\eqref{ratio} can be easily tested on an arbitrary 
geometry by applying a conformal mapping $z'(z)$. Since all the dimensionful parameters in~\eqref{corr} cancel when computing \eqref{ratio},
one has only to express $\eta$ in the new coordinates $z'$. Finally we observe that denoting $1-\eta=\varepsilon$
one obtains~\cite{Suppl}  the small $\varepsilon$ expansion for the ratio in \eqref{ratio}
\begin{equation}
\label{log}
R=1-\varepsilon^{1/2}[\rm{a_0}+\rm{a_1}\varepsilon+
\rm{a_2}\varepsilon^2(1+\rm{b}\log\varepsilon)+O(\varepsilon^3)].
\end{equation}
The logarithmic singularity arises from the mixing of the level 
two descendants of $\phi_{1,3}$ with the field $\phi_{1,5}$ that 
have at $c=1/2$ the same conformal dimension $h_{1,5}=5/2$. This is 
 the first
example where a logarithmic singularity is explicitly calculated
in the context of the critical Ising model.  The logarithmic behaviour in~\eqref{log} has a completely different  origin respect to the well-known logarithmic divergence
of the specific  heat at the critical temperature~\cite{Aldo_rev, Giuseppe_book}.
It shows that the phenomenon of mixing of scaling fields 
and non-diagonalizability of the conformal dilation operator could arise 
potentially at any rational value of the central charge. A circumstance that was already 
recognized in~\cite{Cardy_revLog, VJ_log} and~\cite{VSJ2, GV, SV, KF}
but for which in the Ising model no explicit result was available.  In \cite{SA_1, SA_2}, a
possible source of logarithmic behaviour was also identified but appeared to be ruled out by numerical data.
\paragraph*{Numerical results--}
\begin{figure}
\centering
\includegraphics[width=0.95\columnwidth]{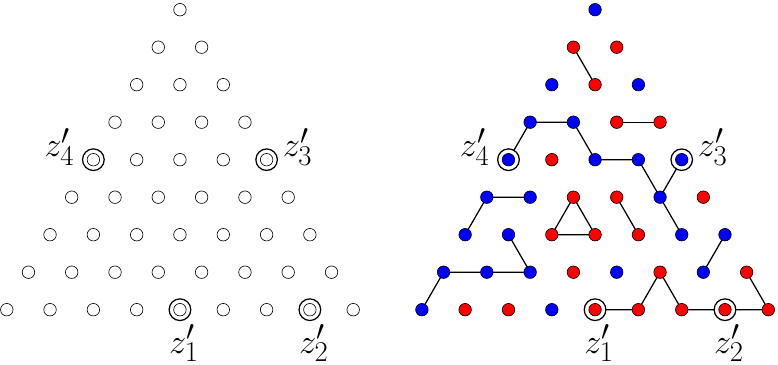}
\caption{(Left) Triangular lattice ($L=9$) with the four points $z'_1$, $z'_2$, $z'_3$ and $z'_4$
highlighted. (Right) a realization of FK clusters contributing
to the probability $P_{(12)(34)}$.}\label{fig1}
\end{figure}

\begin{figure}
\includegraphics[angle=0,width=\columnwidth]{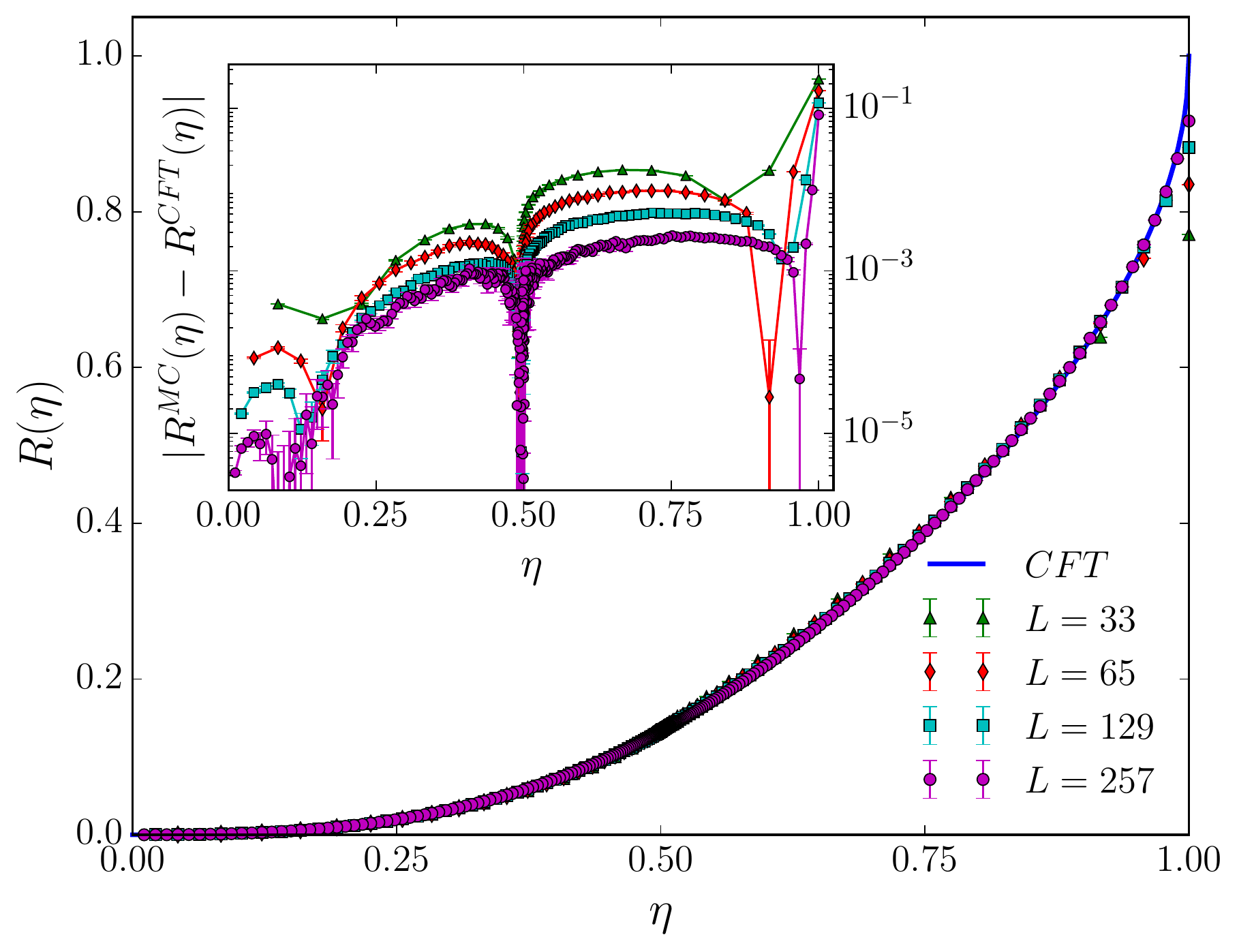}
\caption{Universal ratio $R(\eta)$~\eqref{ratio} for 
the lattice sizes $L=33, 65, 129, 257$ denoted by triangles,
diamonds, squares and circles respectively.  Errors are smaller than the
symbol size. The CFT prediction is plotted with
the continuous line. In the inset deviations of MC data from
the theory are shown with the same symbols used in the
main figure, lines are just guides to the eyes.}\label{fig2}
\end{figure}
Simulations have been carried out on the Ising model at the exactly known critical temperature on a triangular
lattice in triangles of sides of lengths $L=9,17,33,65,129,257$ with open boundary conditions collecting a 
number of samples up $10^{10}$. The random number generator employed is given in~\cite{matsumoto_1998}. 
We implemented the efficient Swendsen-Wang algorithm \cite{swendsen_1987} that provides
direct access to the FK clusters \cite{FK}.

In order to use our results for the upper half plane~\eqref{ratio} in the triangle geometry a Schwarz-Christoffel 
is in order. Given a $z$ in $\mathbb H$ and a $z'$ belonging to the interior of an equilateral 
triangle with vertices $\left(-1, 1, i\sqrt{3}\right)$ the mapping reads $z'=\frac{2 z \Gamma\left(\frac{5}{6}\right) {}_2F_1\left(\frac{1}{2},\frac{2}{3};\frac{3}{2};z^2\right)}
{\sqrt{\pi}\Gamma\left(\frac{1}{3}\right)}$,
${}_2F_1$ being the Gauss hypergeometric function.
In the simulations the three points $z_1'$, $z_3'$ and $z_4'$ have been fixed in the 
midpoint of each side
while the point $z_2'$ takes any position on the boundary between $z'_1$ and $z_3'$. Since the problem
is symmetric under rotation of $2\pi/3$ and $4\pi/3$ around the center of the triangle also the 
configurations obtained with these rotations have been measured to enhance the statistics.
An example of the simulated system together with a realization of FK 
clusters is presented in Figure~\ref{fig1}.
The ratios $P_{(12)(34)}/P_{t}, P_{(14)(23)}/P_{t}, P_{(1234)}/P_{t}$, because of the 
symmetry $P_{(12)(34)}(\eta)=P_{(14)(23)}(1-\eta)$ are not independent and
only one function suffices to specify all of them,  that is $R(\eta)$ as defined
in \eqref{ratio}. In Figure \ref{fig2} we
show the simulation results together with the CFT prediction  for $R$ 
for the four largest lattices. In the inset of Figure~\ref{fig2} we
show the deviations from the exact result.

\paragraph*{Conclusion--}
In this letter we have calculated  the four point connectivities of FK clusters in the critical 
Ising model and show that they can display logarithmic branch cuts. This is a first explicit example 
where such type of singularities 
 are determined exactly for a theory that also has a non-trivial sector belonging to the series of unitary minimal models. Previous exact CFT
studies focused on percolation and SAW~\cite{FSK2, KF_1,KF_2, KF}. 
 Our findings are fully corroborated by numerical simulations, showing excellent agreement. 
 Similar structures, expected in many other important two-dimensional models
including critical percolation, fully deserve the attention of future
investigations. It would be also of clear interest to analyze whether  logarithmic singularities could be found in higher dimensions~\cite{Bootstrap, ABCD}, for 
instance in the three dimensional critical Ising model.

\paragraph*{Acknowledgements.}
GG thanks the International Institute of Physics of Natal for the kind hospitality. JV is grateful to J. Dubail, G. Mussardo and R. Santachiara for a critical
reading of the manuscript and many important and interesting observations. We thank also D. Bernard and A. Cappelli for correspondence and in particular J. Cardy for pointing out his result
in~\cite{Cardy_revLog} about logarithmic observables in the Ising model.

\clearpage


\section*{Supplementary material}

# Supplementary Material
*Exact logarithmic four-point functions in the critical two-dimensional Ising model*

**SOME TECHNICAL DETAILS ON THE SOLUTION OF THE DIFFERENTIAL EQUATION**

We give here an account on how the differential equation for the four point function of the field $\phi_{1,3}$ is solved. The field $\phi_{1,3}$ with scaling dimension $h_{1,3}$ is degenerate and has a null vector at level three. The highest weight state associated to such a field is $|h_{1,3}\rangle = \phi_{1,3}(0)|0\rangle$ and let us denote by $L_n$ the Virasoro algebra generators

$$[L_n, L_m] = (n-m)L_{n+m} + \frac{c}{12}\,\delta_{n+m,0}\,n(n^2-1), \tag{S1}$$

being $c$ the central charge. The null vector $|\chi_{1,3}\rangle$ at level three of $|h_{1,3}\rangle$ can be determined by writing the level three Gram matrix generated by $L_{-1}^3|h_{1,3}\rangle$, $L_{-1}L_{-2}|h_{1,3}\rangle$, $L_{-3}|h_{1,3}\rangle$ and finding its kernel. One finds using the commutation relations (S1)

$$|\chi_{1,3}\rangle = \left[L_{-3} - \frac{2}{h_{1,3}}L_{-1}L_{-2} + \frac{1}{(h_{1,3}+2)(h_{1,3}+1)}L_{-1}^3\right]|h_{1,3}\rangle. \tag{S2}$$

Through the Ward identity, the action of the Virasoro algebra generators on the highest weight state can be converted into the action of differential operators on correlation functions of the primary field $\phi_{1,3}$ [1, 2]. In particular the operator $L_{-n}$ is equivalent to the linear differential operator

$$\mathcal{L}_{-n} = -\sum_{i\neq 1}\frac{(1-n)h_{1,3}}{(z_j-z_i)^n} - \frac{1}{(z_j-z_i)^{n-1}}\frac{\partial}{\partial z_j}, \tag{S3}$$

when acting on the four point function $\langle\phi_{1,3}(z_4)\phi_{1,3}(z_3)\phi_{1,3}(z_2)\phi_{1,3}(z_1)\rangle$. Global conformal invariance fixes the form of this four point function to be

$$\langle\phi_{1,3}(z_4)\phi_{1,3}(z_3)\phi_{1,3}(z_2)\phi_{1,3}(z_1)\rangle = \left[\frac{z_{42}z_{31}}{z_{21}z_{43}z_{32}z_{14}}\right]^{2h_{1,3}} F(\eta); \tag{S4}$$

where $z_{ij} = z_i - z_j$ and $\eta = \frac{z_{21}z_{43}}{z_{42}z_{31}}$ is the harmonic ratio. We can then plug the form (S4) into the differential equation obtained substituting (S3) into (S2). After factoring out the prefactor multiplying the function $F(\eta)$ in (S4) we can take advantage of global conformal invariance setting $z_1 = 0$, $z_3 = 1$ and $z_4 = \infty$; in those limits the harmonic ratio $\eta$ coincides

with the coordinate $z_2$. Notice that when the four points are ordered on the real axis $z_1 < z_2 < z_3 < z_4$, the harmonic ratio $\eta$ can be taken in the domain $0 < \eta < 1$.

At $c = 1/2$ we have $h_{1,3} = 1/2$ and we can then derive the following differential equation for the function $F$

$$[2\eta(1-\eta)]^2 F''' - 3(1-\eta+\eta^2)F' + 3(2\eta-1)F = 0. \tag{S5}$$

For other values of $c$ associated to the critical $Q$-state Potts model a similar equation can be derived, we leave this analysis for future work. It is easy to see that the polynomial $F_{1,1}(\eta) = 1 - \eta + \eta^2$ is a solution of (S5). Actually the same polynomial is a also a solution of the linear differential equation

$$3\eta(1-\eta)F''(\eta) - 2(1-2\eta)F'(\eta) - 2F(\eta) = 0 \tag{S6}$$

that through (S4) analogously expresses the four-point function of the operator $\phi_{2,1}$ at $c = 1/2$. The other solution of (S6) is the Legendre $Q$ function of the second kind $Q_{1/3}^{5/3}(2\eta-1)$ that apparently has not yet a physical interpretation. Coming back to the equation (S5) and substituting $F(\eta) = F_{1,1}(\eta) \int_0^\eta d\eta' G(\eta')$ we can lower the degree obtaining a second order linear differential equation for the function $G(\eta)$

$$3(-1+2\eta+5\eta^2-14\eta^3+7\eta^4)G(\eta)+4(1-\eta)^2\eta^2[3(2\eta-1)G'(\eta)+(1-\eta+\eta^2)G''(\eta)] = 0 \tag{S7}$$

The equation (S7) is a linear differential equation of degree 2 with coefficients that are polynomials with rational coefficients ($\in \mathbb{Q}[\eta]$) and it is of the general form recently analyzed by E. Imamoglu and M. Van Hoeij in [3]. In particular, in [3] the authors developed an algorithm to find (if existing) hypergeometric solutions for this class of differential operators. The algorithm looks for solutions that can be expressed as

$$A(\eta) \, _2F_1(a,b,c,f(\eta)) + B(\eta) \, _2F_1'(a,b,c,f(\eta)), \tag{S8}$$

where $A, B$ are in general algebraic functions, $a, b, c \in \mathbb{Q}$ and the algebraic function $f$ is called the pull-back. One of author made available an implementation for *Maple* of the algorithm that can be freely downloaded at [4]. We ran then the code on the differential equation (S7) finding the two linearly independent [5] solutions $f(\eta)$ and $f(1-\eta)$, where

$$f(\eta) = \frac{16}{21\pi} \frac{((2-\eta)(1+\eta)(-1+2\eta)E(\eta) + (2+\eta(-4+\eta+\eta^2))K(\eta))}{\sqrt{(1-\eta)\eta}(1+(-1+\eta)\eta)^2} \tag{S9}$$







\begin{thebibliography}{99}
\bibitem{note} According to Frobenius ODE theory, the function $F_{1,3}(\eta)$ is logarithmic, threfore $F_{1,5}(1-\eta)$ also has a logarithmic singularity
when $\eta\rightarrow 0$ as we will see.
\bibitem{Foot} Invariant under the exchange of $\eta$ with $1-\eta$, that corresponds to exchange the points $x_1$ with $x_3$.
\end{thebibliography}
\end{document}